\documentclass[twocolumn]{aastex62}

\usepackage{graphicx}
\usepackage{amssymb}
\usepackage{xspace}
\usepackage{outlines}
 \renewcommand{\thefootnote}{\fnsymbol{footnote}}

\newcommand{\Sersic}{S\'{e}rsic }
\newcommand{\galfit}{\textsc{Galfit }}
\newcommand{\dev}{de Vaucouleurs' }
\newcommand{\hst}{\textit{HST} }

\begin{document}
\title{The Structure of Tidal Disruption Event Host Galaxies on Scales of Tens to Thousands of Parsecs} 

\author[0000-0002-4235-7337]{K. Decker French}
\altaffiliation{Hubble Fellow}
\affil{Observatories of the Carnegie Institute for Science, 813 Santa Barbara St. Pasadena, CA 91001, USA}
\author[0000-0001-7090-4898]{Iair Arcavi}
\affil{School of Physics and Astronomy, Tel Aviv University, Tel Aviv 69978, Israel}
\affil{CIFAR Azrieli Global Scholars program, CIFAR, Toronto, Canada}
\author[0000-0001-6047-8469]{Ann I. Zabludoff}
\affil{Steward Observatory, University of Arizona, 933 N Cherry Ave, Tucson, AZ 85721, USA}
\author[0000-0002-4337-9458]{Nicholas Stone}
\affil{Racah Institute of Physics, The Hebrew University, Jerusalem 91904, Israel}
\author[0000-0002-1125-9187]{Daichi Hiramatsu}
\affil{Department of Physics, University of California, Santa Barbara, CA 93106-9530, USA}
\affil{Las Cumbres Observatory, 6740 Cortona Drive, Suite 102, Goleta, CA 93117-5575, USA}
\author[0000-0002-3859-8074]{Sjoert van Velzen}
\affil{Center for Cosmology and Particle Physics, New York University, New York, NY 10003, USA}
\affil{Department of Astronomy, University of Maryland, College Park, MD 20742, USA}
\author[0000-0001-5807-7893]{Curtis McCully}
\affil{Department of Physics, University of California, Santa Barbara, CA 93106-9530, USA}
\affil{Las Cumbres Observatory, 6740 Cortona Drive, Suite 102, Goleta, CA 93117-5575, USA}
\author[0000-0002-7152-3621]{Ning Jiang}
\affil{Key Laboratory for Research in Galaxies and Cosmology, Department of Astronomy, University of Science and Technology of China, Chinese Academy of Sciences, Hefei, Anhui 230026, People's Republic of China}
\affil{School of Astronomy and Space Sciences, University of Science and Technology of China, Hefei, Anhui 230026, People's Republic of China}

\begin{abstract}

We explore the galaxy structure of four tidal disruption event (TDE) host galaxies on 30 pc to kpc scales using \hst WFC3 multi-band imaging. The star formation histories of these hosts are diverse, including one post-starburst galaxy (ASASSN-14li), two hosts with recent weak starbursts (ASASSN-14ae and iPTF15af), and one early type (PTF09ge). Compared to early type galaxies of similar stellar masses, the TDE hosts have higher central surface brightnesses and stellar mass surface densities on 30--100 pc scales. The TDE hosts do not show the large, kpc-scale tidal disruptions seen in some post-starburst galaxies; the hosts have low morphological asymmetries similar to those of early type galaxies. The lack of strong asymmetries are inconsistent with a recent major ($\sim$1:1 mass) merger, although minor ($\lesssim$1:3) mergers are possible. Given the time elapsed since the end of the starbursts in the three post-burst TDE hosts and the constraints on the merger mass ratios, it is unlikely that a bound supermassive black hole binary (SMBHB) has had time to coalesce. The TDE hosts have low central ($<140$ pc) ellipticities compared to early type galaxies. The low central ellipticities disfavor a strong radial anisotropy as the cause for the enhanced TDE rate, although we cannot rule out eccentric disks at the scale of the black hole gravitational radius of influence ($\sim1$ pc). These observations suggest that the high central stellar densities are a more important driver than SMBHBs or radial anisotropies in increasing the TDE rate in galaxies with recent starbursts.

\end{abstract}

\section{Introduction}
Tidal Disruption Events (TDEs) are observed when a star passes close enough to a supermassive black hole; the strong tidal forces of the supermassive black hole are capable of disrupting the star and accreting some of the disrupted star's material \citep[e.g.,][]{Hills1975,Rees1988}.

The rate of optical/UV-bright and X-ray bright TDEs has been observed to depend on the recent global star formation history of the host galaxy \citep{Arcavi2014}, with TDE rates enhanced by $\sim20-200\times$ in quiescent Balmer-strong and post-starburst galaxies \citep{French2016, Law-Smith2017, Graur2018}. This connection between the small-scale dynamics affecting the nucleus and the large-scale dynamics affecting the total star formation history of the galaxy presents a puzzle: what connects these two very different scales? Several intriguing possibilities have been presented. 

One possibility is that post-starburst galaxies have high central concentrations of stars, which leads to a greater number of stars able to populate the loss cone of stars which can be tidally disrupted. Indeed, there is evidence that TDE host galaxies have higher central concentrations on kpc-scales from SDSS imaging \citep{Law-Smith2017, Graur2018}, and we expect post-starburst galaxies to have centrally-concentrated stellar distributions caused by merger-triggered bursts of star formation \citep[e.g.,][]{Yang2008}. In this work, we aim to test the central stellar densities of the TDE host galaxies down to spatial scales of 30--100 pc using high resolution \hst imaging. While these spatial scales are still higher than the gravitational radius of influence of the black hole ($\sim 0.3-3$ pc for the TDE host galaxies targeted here), \hst observations of the nearest TDE host galaxies are the best probe of the central densities of these galaxies, and these scales are within the mean break radius of early type galaxy light profiles, so they are likely to be correlated with densities in smaller spatial scales. 

Another possibility for explaining the high TDE rate in post-starburst and quiescent Balmer-strong galaxies is if the galaxies are in a post-merger phase where the supermassive black hole binary (SMBHB) is inspiralling. SMBHBs can increase the TDE rate by a large amount over a short period of time \citep[e.g.,][]{Chen2011}. Mergers can both produce a SMBHB and trigger a starburst, so establishing whether the TDE host galaxies have experienced a recent merger (major or minor) is important for assessing whether the TDE rate is impacted by SMBHB effects. \hst imaging sensitive to faint tidal features can determine whether the TDE host galaxies have had a recent (in the last $\sim$Gyr) major merger. 

Several dynamical effects have also been proposed to account for the high TDE rate; a radial anisotropy \citep{Stone2018} or eccentric disk \citep{Madigan2018} could increase the TDE rate, and may be expected to occur in the Gyr after a galaxy-galaxy merger, during the post-starburst phase. In the case of a large radial anisotropy, tidally disrupted stars could originate from well outside the influence radius ($\sim10-30$ pc), and the ellipticity of the stellar light at these scales can constrain the dynamical possibilities for enhancing the TDE rate. Another possibility for increasing the TDE rate after a merger is the scenario described by \citet{Cen2020}, where a massive $\sim100$ pc disk forms around the primary SMBH, producing a high rate of TDEs around the secondary SMBH as it inspirals through the disk. These spatial scales are resolvable for the most nearby TDE host galaxies with \hst.

In this work, we aim to test these effects using high resolution \hst imaging of four TDE host galaxies. These data allow us to probe the central concentrations and eccentricities down to the $<100$ pc scale, search for faint post-merger tidal features, and compare the TDE hosts to detailed \hst studies of post-starburst and early type galaxies. We describe the data and reduction in \S2, the results of our morphological analyses in \S3, discuss the implications of our findings on the high TDE rate in post-starburst galaxies in \S4, and summarize and conclude in \S5.

\section{Observations and Data Reduction}
\subsection{TDE Host Galaxy Sample}
The imaging targets were selected to be the four nearest TDE host galaxies from the sample of eight UV/optical bright TDEs with broad H/He lines from \citet{French2016}: ASASSN-14li \citep{Holoien2015}, ASASSN-14ae \citep{Holoien2014}, iPTF15af \citep{Blagorodnova2019}, and PTF09ge \citep{Arcavi2014}. Of the four TDE host galaxies studied here, one (ASASSN-14li) is a post-starburst, two (ASASSN-14ae and iPTF15af) are quiescent Balmer-strong, and one (PTF09ge) is a quiescent early type according to their spectra \citep{French2016}. PTF09ge however has star formation missed by the 0.9 kpc aperture spectrum in its outskirts, revealed by the blue ring in our imaging (Figure \ref{fig:color}). Our stellar population fitting analysis also revealed a recent short burst of star formation in this galaxy, similar to the star formation history of iPTF15af, although this analysis used the same small aperture spectrum \citep{French2017}. 

This sample is representative of the host galaxy types of known TDEs. Using the sample of TDEs with host galaxy data compiled in French et al. (2019 ISSI review, submitted; updated from \citealt{Graur2018}), of the optical/UV-bright TDEs with broad H/He lines (like those identified by \citet{Arcavi2014}), 60\% are quiescent Balmer strong (including post-starburst galaxies), and 33\% are post-starburst. Of the X-ray TDEs with host galaxy data (including those designated by \citet{Auchettl2017} as ``likely" or ``possible" TDEs), 32\% are quiescent Balmer-strong and 12\% are post-starburst. For the more robust subset of X-ray TDEs, 75\% are quiescent Balmer-strong and 25\% are post-starburst.  The post-starburst host of ASASSN-14li is unique among the four galaxies studied here in being the only post-starburst galaxy, but having a sample with 1/4 post-starburst galaxies is consistent with post-starburst galaxies comprising 12\%-33\% of the TDE host population. Similarly, the fraction of quiescent/early-type and quiescent Balmer-strong galaxies in the {\it HST}--observed sample is representative of the known TDE host population, spanning the range of likely recent star formation histories.

Given the black hole masses measured by \citet{Wevers2017} using the bulge velocity dispersions and by \citet{Mockler2018} using the TDE light curves, and the fitting functions from \citet{Stone2016b} for all and cuspy galaxies, the radius of influence of the SMBH is 1--3 pc, 0.3--0.9 pc, 2.4--2.7 pc and 1.0-1.6 pc for the host galaxies of ASASSN-14li, ASASSN-14ae, iPTF15af, and PTF09ge respectively. 

\subsection{\hst WFC3 Imaging}
Each of the four galaxies was observed using three bands: F438W, F625W, and F814W. The two bluer bands were selected to match those used by \citet{Yang2008} in their study of post-starburst galaxies. The F814W band was then added to best measure the stellar bulge properties. Each galaxy was observed to a depth of $\sim$26 mag/arcsec$^2$ in order to measure faint tidal features. The observations were carried out in 2/2017--8/2017 as part of program 14717 (PI: Arcavi) during Cycle 24. Details of the observations are shown in Table \ref{tab:obs}.

We use the flat fielded, charge transfer efficiency corrected, {\tt *.flc} files from the \hst archive. The individual exposures are then re-aligned and stacked using SNHST \citep{mccully2018}. This procedure uses PanSTARRS imaging \citep{Chambers2016} to refine the astrometric solution to best match the images from multiple visits to each object. The images are then combined using DrizzlePac\footnote{\url{http://www.stsci.edu/scientific-community/software/drizzlepac}}. Cosmic rays are removed using AstroScrappy \citep{mccully2018a}.

\begin{figure*}
\includegraphics[width = 0.5\textwidth]{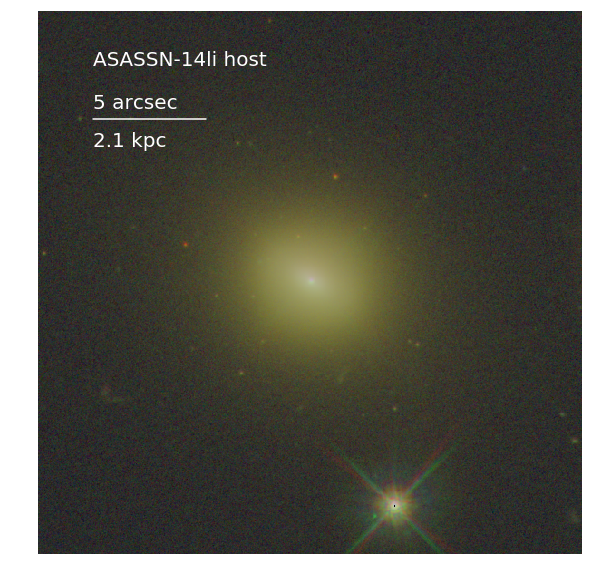}
\includegraphics[width = 0.5\textwidth]{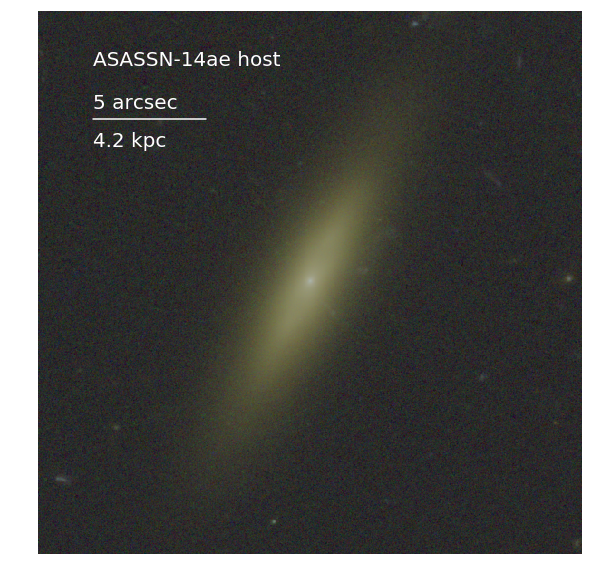}
\includegraphics[width = 0.5\textwidth]{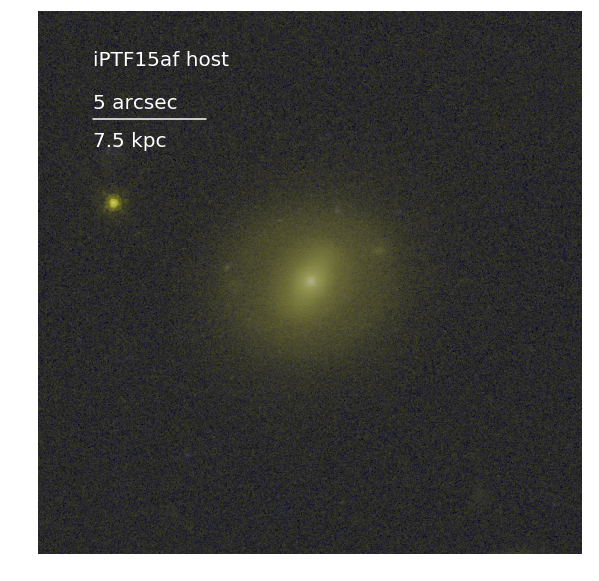}
\includegraphics[width = 0.5\textwidth]{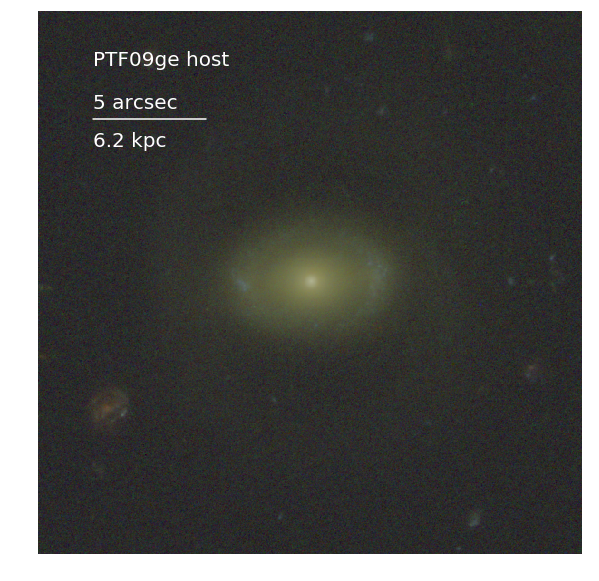}
\caption{Three-color images (F814W, F625W, F438W) of the four TDE host galaxies targeted here. The TDE host galaxies display a wide range of morphologies. The host galaxy of ASASSN-14li has a bright central point source that can be seen through its diffraction spikes in the optical images. The host galaxy of ASASSN-14ae is an edge-on disk that also has a bright central bulge. The host galaxies of iPTF15af and PTF09ge are barred lenticular galaxies; both have blue disk components in their outskirts in addition to the bright central bulges. The host of iPTF15af has a series of blue clumps in its outer disk above the galaxy (to the north). The host of PTF09ge has extended arms reaching from the two blue knots in the disk. These can be better seen in the higher contrast stretch images in Figure \ref{fig:stretch}. Two of the host galaxies (iPTF15af and PTF09ge) appear to have bar-like structures, consistent with $\sim40$\% of low redshift galaxies having bars \citep{Elmegreen2004}. While the host galaxies span a range of properties, we note a lack of grand design spirals or irregular dwarf galaxies.
}
\label{fig:color}
\end{figure*}

The resulting 3-color images can be seen in Figure \ref{fig:color}. In Figure \ref{fig:stretch} we present high stretch images of the four TDE host galaxies in each of the three bands, optimized to search for low surface brightness tidal features.

\begin{table*}[]
    \centering
    \begin{tabular}{lllllll}
    \hline
    TDE & R.A. & decl. & z & Pixel size (pc)$^a$ & Ref & Peak  \\
    \hline
         ASASSN-14li & 12:48:15.23 &	+17:46:26.44 & 0.02058 & 17 & \citep{Holoien2015}  & 2014-11-14$^b$\\
         ASASSN-14ae & 11:08:40.12	& +34:05:52.23 & 0.0436 & 33 & \citep{Holoien2014} & 2014-01-25$^b$ \\
         iPTF15af  & 08:48:28.12 &  +22:03:33.58 & 0.079 & 60 & \citep{Blagorodnova2019} & 2015-02-23 \\
         PTF09ge & 14:57:03.18	& +49:36:40.97 & 0.064 & 50 & \citep{Arcavi2014} & 2009-06-13\\
         \hline
    \end{tabular}
    \caption{Observed TDE Targets. $^a$ physical size probed by 0.04\arcsec\ pixels. $^b$ Discovery date, discovered after peak. }
    \label{tab:targets}
\end{table*}

\begin{table}[]
    \centering
    \begin{tabular}{lll}
    \hline
    Image & Exp. Time (s) & Start Time \\
    \hline
         ASASSN-14li host \\
         F814W & 2496 & 2017-05-31 11:07:49 \\
         F625W & 2496 & 2017-05-31 09:32:28 \\
         F438W & 2476 & 2017-05-19 21:18:49 \\
         \hline
         ASASSN-14ae host \\
         F814W & 2520 & 2017-03-08 06:37:21 \\
         F625W & 2520 & 2017-04-05 20:02:51 \\
         F438W & 5662 & 2017-05-14 12:37:04 \\
         \hline
         iPTF15af host \\
         F814W & 2496 & 2017-02-25 16:14:50 \\
         F625W & 5616 & 2017-02-21 15:17:26 \\
         F438W & 8480 & 2017-02-23 11:47:44 \\
         \hline
         PTF09ge host \\
         F814W & 2624 & 2017-03-12 00:18:54 \\
         F625W & 2624 & 2017-03-12 22:19:20 \\
         F438W & 5876 & 2017-08-04 00:51:57 \\
         \hline
    \end{tabular}
    \caption{Description of Observations}
    \label{tab:obs}
\end{table}

\subsection{PSF fitting}
\label{sec:psf}
We explore three possible solutions for determining the point spread function (PSF) for each stacked galaxy image in each band. (1) The \hst PSF library contains PSF profiles for the F438W and F814W bands\footnote{\url{http://www.stsci.edu/hst/instrumentation/wfc3/data-analysis/psf}}. (2) We also consider the PSFs calculated using Tiny Tim\footnote{\url{http://tinytim.stsci.edu/cgi-bin/tinytimweb.cgi}} \citep{Krist2011}. With Tiny Tim, we assume a A5V star, similar to the stellar population dominating the spectra of these galaxies, and use the central position of the galaxy on the chip. (3) Three of the four galaxies (hosts of ASASSN-14li, iPTF15af, and PTF09ge) are in fields with suitable foreground stars to calculate PSFs, that are bright enough to have high SNR, but not saturated. To calculate the best-fit PSFs from stars in each field (except for the ASASSN-14ae field), we first mask the diffraction spikes, then iteratively fit a series of Gaussian profiles using \galfit \citep{Peng2002, Peng2010}. We follow \citet{Stone2016} and add Gaussian components until the $\chi^2/\nu$ value stops decreasing. For our sample of images, between 4-5 Gaussian components are chosen. 

\begin{figure}
\centering
\includegraphics[width = 0.4\textwidth]{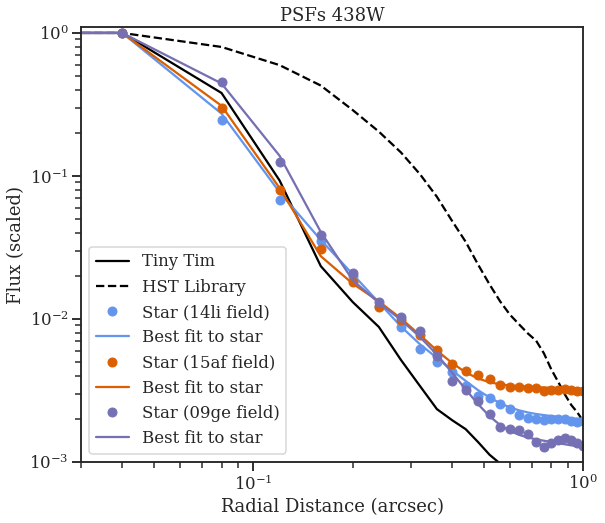}
\includegraphics[width = 0.4\textwidth]{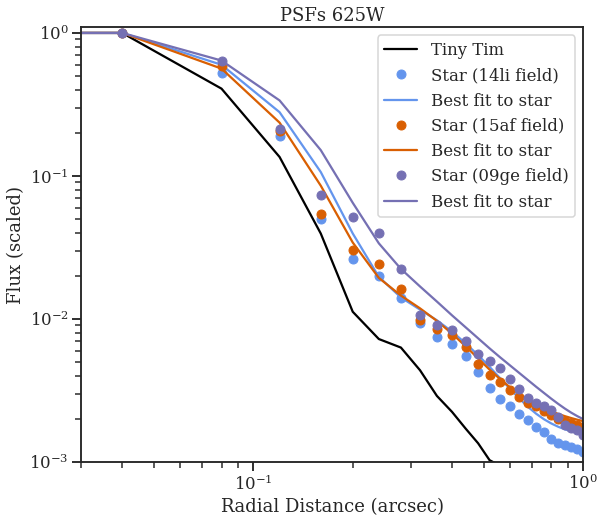}
\includegraphics[width = 0.4\textwidth]{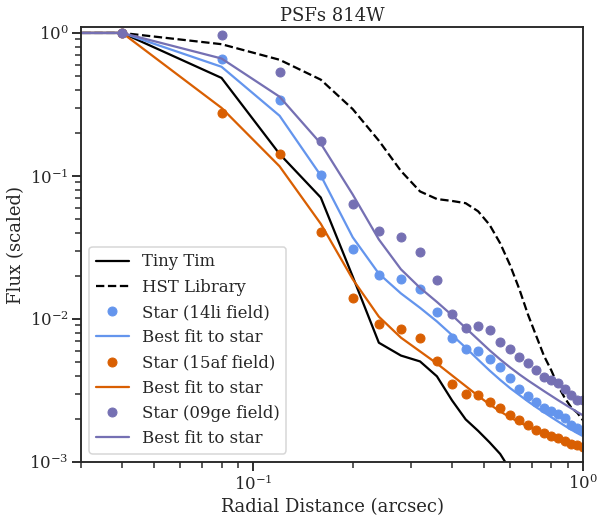}
\caption{PSFs in each band from Tiny Tim, the \hst PSF library, and fits to stars in the three fields with suitable bright stars. We note that the library and Tiny Tim PSFs have not been corrected for the drizzle or co-add steps applied to the science images. The best-fit PSFs generally lie between the Tiny Tim and Library PSFs, and are consistent between fields except for the F814W
filter images. We adopt the best-fit star PSF specific to that field in each case, and use the iPTF 15af field star for the ASASSN 14ae analysis. }
\label{fig:psfs}
\end{figure}

The results of the three methods for determining the PSF are shown in Figure \ref{fig:psfs} for each band. We find that the library PSFs are significantly wider than the other measures. The Tiny Tim PSFs are often narrower than the profiles and best-fit models to stars in each frame. We note that the library and Tiny Tim PSFs have not been corrected for the drizzle or co-add steps applied to the science images. For F438W and F625W, the stellar PSFs are very similar for each field, but we observe significant differences for the F814W fields. We adopt the best-fit stellar PSF for each field--band combination. For the ASASSN-14ae field, we test for the best PSF from the other three fields by fitting a \Sersic profile to the ASASSN-14ae host galaxy using each of the 3 PSFs, and choosing the one with the lowest $\chi^2/\nu$. The best PSF found using this method is from the iPTF15af field, so we adopt this PSF for all of the fits to the ASASSN-14ae host galaxy presented here.

\section{Results}

\subsection{TDE Host Galaxy Morphologies}

From Figure \ref{fig:color}, it is clear that the TDE host galaxies have a range of morphologies, although none are grand design spirals or irregular dwarf galaxies. The host galaxy of ASASSN-14li has a bright central point source that can be seen through its diffraction spikes in the optical image. The host galaxy of ASASSN-14ae is an edge-on disk that also has a bright central bulge. The host galaxies of iPTF15af and PTF09ge are barred lenticular galaxies; both have blue disk components in their outskirts in addition to bright central bulges. The host of iPTF15af has a series of blue clumps in its outer disk to the north. These clumps may be star-forming knots, dwarf galaxies, minor merger remnants, or chance projections along the line of sight. The host of PTF09ge has extended arms reaching from the two blue knots in the disk. The symmetry and blue color of the extended arms indicates they may be spiral arms, but such features could also be caused by a recent merger. These extended arms also closely resemble the ring-like ``$R_1^\prime$" morphology described by \citet{Buta2017} (see e.g., UGC 12646 in \citealt{Buta2017} Figure 1), and do not require mergers or other non-secular processes to form. Bar-like structures are seen in two of the host galaxies (iPTF15af and PTF09ge), consistent with $\sim40$\% of low redshift galaxies having bars \citep{Elmegreen2004}. 

No significant dust lanes are seen in the four TDE hosts, aside from the dust seen in the spiral arms around the PTF09ge host galaxy. This is consistent given Poisson errors with the rate of 7/21 seen by \citet{Yang2008} in post-starburst galaxies. Of the seven post-starburst galaxies with dust lanes, \citet{Yang2008} find three instances where the dust lanes cross the bulge of the galaxy. It is not surprising to find no cases of central dust lanes in the TDE host galaxy sample, given the difficulty of detecting a TDE through significant dust attenuation, but a larger sample would be needed to establish a statistically significant difference from the post-starburst sample.

We discuss the galaxy morphologies in more quantitative detail in the remainder of this section by fitting their two-dimensional surface brightness distributions, examining their color gradients, and searching for asymmetric features.

\subsection{Surface Brightness Profile Fitting}
\label{sec:sb_profiles}
We fit the surface brightness profile of each galaxy in each band using \galfit \citep{Peng2002, Peng2010}. We use the PSF models derived from stars in each field as described in \S\ref{sec:psf}. We fit a sky level with no gradient across the field in addition to the models discussed below. For galaxies with nearby projected companions, we fit separate \galfit model components. We fit separate \galfit components for the projected nearby edge-on disk galaxy and foreground star near the host galaxy of ASASSN-14li. We fit a separate component for the foreground star projected near the host galaxy of iPTF15af, and the foreground star projected near the host galaxy of PTF09ge.

We consider six\footnote{We also attempted to fit Nuker profiles \citep{Lauer1995}, but these failed to converge in most cases and were highly dependent on the initial parameter guesses. In the cases that did converge, these fits were never better than the six described in this section. This is likely because of the limited resolution inside the break radius.} types of models for each galaxy in each band: (1) a \Sersic profile, with $n$ free to vary, (2) a \Sersic profile plus a central point source, (3) a \Sersic profile plus an exponential disk, (4) a \dev profile, (5) a \dev profile plus a central point source, and (6) a \dev profile plus a disk. \Sersic profiles \citep{Sersic1963, Sersic1968} are parameterized using an index such that higher \Sersic indices are more centrally concentrated. A \Sersic index of $n=1$ corresponds to an exponential disk and fits typical star-forming disk galaxies, and higher \Sersic indices of $n\sim4$ (with the special case of $n=4$ being the \dev profile, \citealt{dev1948}) are good models for early type galaxies. Using \galfit to model combinations of models helps fit galaxies with multiple bulge plus disk components. Post-starburst galaxies are often best-fit with high \Sersic indices $n>4$ \citep{Yang2008} and often have central point sources \citep{Yang2006}. Given the long duration of TDE emission (especially in the UV; \citealt{vanvelzen2019}), it is also important to consider possible nuclear point sources if the TDE was still bright 5--8 years after its peak. 

We allow the \galfit models to vary in two-dimensions, fitting parameters for axis ratio and position angle in addition to the radial profile parameters. We select the best-fit model by choosing the one with the lowest $\chi^2/\nu$. 

The best-fit models are shown in Table \ref{tab:galfit} in addition to reduced $\chi^2/\nu$ values and selected parameters from the best-fit models. The host galaxy of ASASSN-14li is best fit by a \Sersic profile plus a central point source for all three bands, and the PTF09ge host galaxy is also best fit by this model for two bands. Both ASASSN-14ae and iPTF15af are best fit by models with disks added for 2--3 of the bands considered here. The best $\chi^2/\nu$ values are typically between 1--2, except for the fits to the PTF09ge host galaxy, which are higher. This indicates the presence of additional un-modeled structure, which we explore further in the subsequent sections. 

The magnitude of the central point source component is 18--19 mag for the cases where the galaxy is best fit with a central point source, and 19--23 mag for cases where a central point source did not improve the fit. These point sources are significantly brighter than we expect the residual TDE emission to be. The brightest residual TDE emission is expected from ASASSN-14li, which would have been 19.1 mag in the NUV at the time of the \hst imaging \citep{vanvelzen2019}. The $g$ band magnitude is expected to be 1.5 mag fainter, $\sim20.6$ mag, which is much fainter than the 18.0--18.5 mag optical point source in the best-fit \galfit model. Thus, the central point sources from the TDE host galaxies are likely persistent, and un-related to the TDE. They may be central star clusters, AGN, or very cuspy stellar density profiles. High resolution spectroscopy will be required to differentiate among these possibilities. 

We see no evidence for dual nuclei \citep[e.g.,][]{Lauer1993} in the stellar light down to the resolution limit of 0.04 arcsec for any of the TDE host galaxies considered here.  The resolution of $\sim$10s of pc constrains any SMBHBs to be smaller than the stage where dynamical friction is affecting their evolution and inspiral \citep{Yu2002,Kelley2017}. The spatial separation where the TDE rate is expected to be enhanced from SMBHB effects is of order the gravitational radius of influence of the primary black hole $r_{\rm infl} \sim 0.3-3$ pc, smaller than what can be constrained with these data \citep{Chen2011,Kelley2017}. We explore the constraints we can place on SMBHB effects using larger $\sim$kpc scale merger features in \S\ref{sec:asym} and \S\ref{sec:smbhb}.

\begin{table*}[]
    \centering
    \begin{tabular}{llcccc}
    \hline
         Image & Best-fit model & $\chi^2/\nu$  & $n$ & $n$ & Point source mag \\
          & & (best-fit model) & (S\'{e}rsic--only fit$^b$) & (\Sersic + ps fit$^b$) & (\Sersic + ps fit$^b$) \\
         \hline
         ASASSN-14li \\
         814W & \Sersic+ ps$^c$ & 1.596 & 5.06 & 3.03 & 18.01 \\
         625W & \Sersic+ ps & 1.610 & 5.09 & 3.21 & 18.40 \\
         438W & \Sersic+ ps & 1.214 & 10.82 & 3.12 & 18.49 \\
         \hline
         ASASSN-14ae \\
         814W & \Sersic+ disk & 1.467 & 2.61 & 1.90 & 20.37 \\
         625W & \Sersic+ disk & 1.422 & 3.11 & 1.77 & 19.86 \\
         438W & \Sersic+ disk & 1.294 & 6.35 & 2.30 & 21.10 \\
         \hline
         iPTF15af \\
         814W & \dev+ disk & 1.324 & 5.72 & 3.72 & 21.12 \\
         625W & \Sersic+ disk & 1.471 & 7.36 & 3.10 & 19.97 \\
         438W & \dev+ ps & 1.227 & 16.21 & 4.61 & 22.44$^a$ \\
         \hline
         PTF09ge \\
         814W & \Sersic+ ps & 3.631 & 10.42 & 2.98 & 19.37 \\
         625W & \Sersic+ ps & 4.140 & 11.99 & 2.82 & 19.77 \\
         438W & \dev+ disk & 6.194 & 14.96 & 5.61 & 22.13 \\
         \hline         
         \hline
    \end{tabular}
    \caption{Best fit \galfit models and \galfit model parameters. $^a$ The point source magnitude for the better-fit \dev+psf model is 22.36 mag. $^b$ These parameter values are the best-fit parameters for these classes of models, even in the cases where this class of model is not the best-fit overall. $^c$ ps = Central point source component. }
    \label{tab:galfit}
\end{table*}

\subsection{Surface Brightness Profile Comparisons to Other Galaxy Types}

\begin{figure*}
\includegraphics[width = 0.49\textwidth]{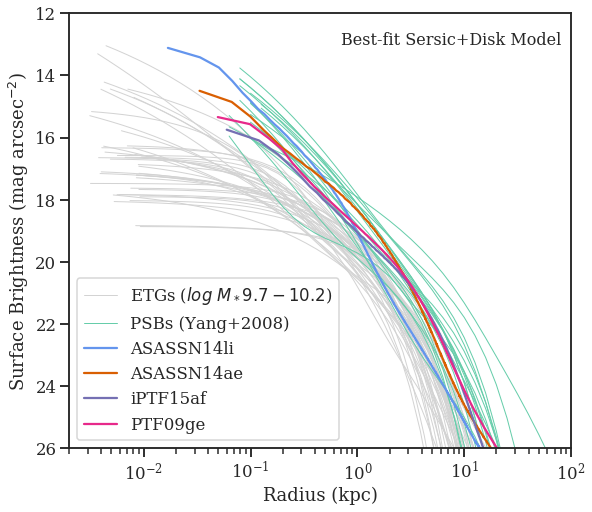}
\includegraphics[width = 0.49\textwidth]{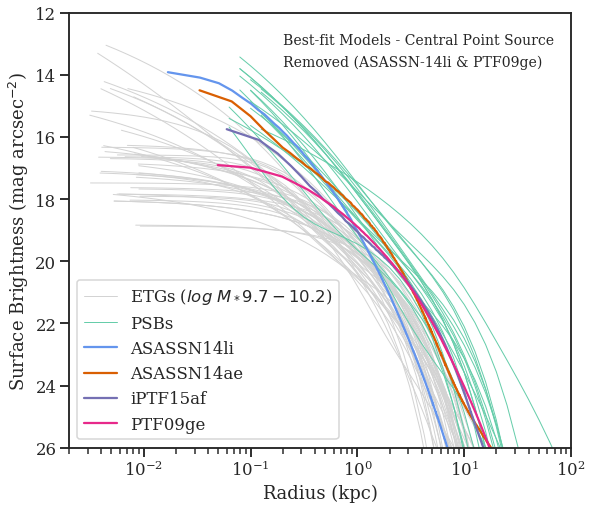}
\caption{\textit{Left:} Radial surface brightness profiles for the four TDE hosts targeted here, as well as comparison early type galaxies \citep{Krajnovic2013} and post-starburst galaxies \citep{Yang2008}. We restrict the stellar masses of the early type galaxy sample to be within the range of the TDE hosts. The \citet{Yang2008} post-starburst fits are \dev plus exponential disk fits. The early type galaxy fits are \Sersic plus disk fits. The TDE host fits are the best-fit \Sersic plus disk models. All models for all samples are PSF-deconvolved. The TDE host galaxies have high central surface brighnesses on the central 30--100 pc scales, similar to the post-starburst galaxies and the brightest high-\Sersic\-index early type galaxies. \textit{Right:} Because two of the TDE hosts (of ASASSN-14li and PTF09ge) are better fit by models including a central point source, we show here the surface brightness profiles for these two host galaxies with the central point source removed. If the central point source component is non-stellar in origin, the host galaxy of ASASSN-14li would still have a bright central surface brightness, similar to post-starburst galaxies, while the host galaxy of PTF09ge would be more similar to early type galaxies at similar stellar masses.
}
\label{fig:profiles}
\end{figure*}

We compare the surface brightness profiles measured above for the TDE hosts to post-starburst galaxies and early type galaxies in Figure \ref{fig:profiles}. Surface brightness profiles for a sample of 260 early type galaxies from the Atlas-3D survey were measured by \citet{Krajnovic2013}. These profiles were modeled using two component \Sersic plus exponential disk models. Surface brightness profiles for 21 post-starburst galaxies were measured by \citet{Yang2008} using two component \dev plus exponential disk models. We correct all surface brightness profiles for cosmological surface brightness dimming. In order to best compare to the $r$-band (peak wavelength $\sim 6231$\AA) data used by \citet{Krajnovic2013}, we use the \hst F625W band imaging of the TDE hosts and post-starbursts for this comparison. We note that the \citet{Yang2008} sample of post-starburst galaxies is at slightly higher redshift ($\langle z \rangle = 0.09$ vs. $\langle z \rangle = 0.05$) than the TDE host galaxies considered here, but the portions of the spectra probed are similar. 

While we can compare the TDE host galaxies to detailed studies of post-starburst and early type galaxies, unfortunately there are no similar analyses of quiescent Balmer-strong galaxies in the literature to which to compare ASASSN-14ae and iPTF15af. Compared to galaxies satisfying a stricter post-starburst definition cut on Balmer absorption, quiescent Balmer-strong galaxies could have either longer post-burst ages, lower fractions of stellar mass created in the recent burst, or longer-duration bursts \citep{French2018b}. However, smaller bursts are more common in galaxies than larger bursts creating more stars, so we found in \citet{French2017} that most quiescent Balmer-strong galaxies have smaller bursts than post-starburst galaxies, with similar ranges of post-burst ages and burst durations. We thus expect them to have properties in between those of post-starburst and early type galaxies, due to the less massive population of recently formed (A) stars\footnote{One concern in making this comparison between the host galaxy types is if it is more difficult to detect nuclear transients in galaxies with bright centers, but the peak optical absolute magnitudes of the known TDEs hosted by quiescent Balmer-strong and post-starburst galaxies are similar.}.

Both the TDE host galaxies and post-starburst galaxies have bright, centrally concentrated stellar light profiles. The TDE host galaxy stellar light profiles stay high down to scales of $<100$ pc and even to $\sim30$ pc for the host galaxy of ASASSN-14li. The TDE host galaxies have central surface brightnesses comparable to the brightest early type galaxies with similar stellar mass, and comparable to the dimmer half of the post-starburst galaxies. Even the TDE host galaxies that are not post-starbursts nonetheless have bright central surface brightnesses more typical of post-starburst galaxies. 

To consider the case where the central point source components of the host galaxies of ASASSN-14li and PTF09ge may be non-stellar, we remove the point source component from the models in the comparison in the right hand panel of Figure \ref{fig:profiles}. With the central point source component removed, the host galaxy of ASASSN-14li would still have a bright central surface brightness, similar to post-starburst galaxies, while the host galaxy of PTF09ge would be more similar to early type galaxies at similar stellar masses.

\begin{figure}
    \centering
    \includegraphics[width = 0.5\textwidth]{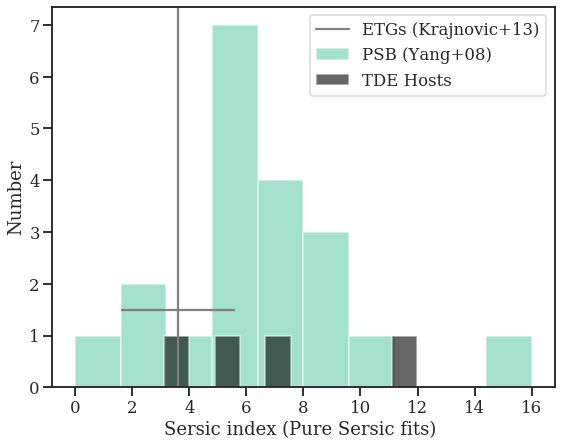}
    \caption{Large (kpc) scale galaxy concentrations parameterized using \Sersic indices from S\'{e}rsic--only \galfit modelling for the TDE host galaxies (grey) and comparison measurements for post-starburst galaxies (green) from \citet{Yang2008}. We consider only the F625W measurements to match \citet{Yang2008}. The two samples cannot be distinguished statistically. Both samples have concentrated profiles with most \Sersic indices $n>4$, higher than typical early type galaxies, even though only one TDE host considered here is a post-starburst galaxy. \citet{Krajnovic2013} measured the mean S\'{e}rsic--only fit \Sersic index for the Atlas-3D sample of early type galaxies to be $\langle n \rangle = 3.6$, with a standard deviation of 2.0 (light grey line and 68 percentile bar). The only TDE host galaxy which is not a quiescent Balmer-strong galaxy, the host of PTF09ge, has the highest \Sersic index of the TDE host galaxies measured using this method. }
    \label{fig:sersic_comp}
\end{figure}

The surface brightness profiles of the TDE host galaxies appear to be more concentrated than early type galaxies of similar stellar mass, so we compare their \Sersic indices as a parameterization of their concentration on larger $\gtrsim$ kpc scales. We found in the previous section that the S\'{e}rsic--only \galfit fits to the TDE host galaxies have high \Sersic indices. While the S\'{e}rsic--only fits are never the best fit of those considered, we use the \Sersic indices measured from these fits to compare to the \Sersic indices measured for a sample of post-starburst galaxies from \citet{Yang2008}, as a way to quantitatively compare the typical surface brightness slopes. This parameterization also allows us to compare the TDE hosts to other samples of galaxies, as the variable \Sersic parameter ensures a range of galaxy morphologies from bulge to disk dominated can be fit with similar accuracy. We plot histograms of the \Sersic indices for each sample in Figure \ref{fig:sersic_comp}. We again use the measurements from the F625W \hst imaging for the TDE host galaxies and post-starbursts and the $r$ band imaging for the early types. 

We compare the \Sersic indices of the TDE host galaxies to measurements of post-starburst galaxies from \citet{Yang2008} and early type galaxies from \citet{Krajnovic2013}. \citet{Krajnovic2013} measured the mean S\'{e}rsic--only fit \Sersic index for the Atlas-3D sample to be $\langle n \rangle = 3.6$, with a standard deviation of 2.0. Most post-starburst galaxies have higher \Sersic indices $n>4$. The TDE host galaxies have high \Sersic indices, similar to the post-starburst sample and higher than typical early type galaxies. The TDE host galaxy and post-starburst galaxy distributions cannot be distinguished statistically using a Kolmogorov-Smirnov test.

The host galaxy of ASASSN-14li has a \Sersic index close to the typical value for post-starburst galaxies, and a bright central surface brightness profile also consistent with this sample. The host galaxy of iPTF15af also has a high \Sersic index, higher than typical early type galaxies and consistent with post-starbursts; the other quiescent Balmer-strong host of ASASSN-14ae has a \Sersic index consistent with early type galaxies. Both hosts have central surface brightnesses similar to the lower half of the post-starburst galaxies and higher than most of the early type galaxies. This is not inconsistent with what we might expect for quiescent Balmer-strong galaxies. The host of PTF09ge has the highest \Sersic index, significantly higher than early type galaxies, and higher than most of the post-starbursts. Given that most star forming galaxies have \Sersic indices of $\sim1$ (exponential disks), the high \Sersic index is unexpected considering the star forming ring.

In several cases, the best-fit \galfit model indicates the high \Sersic indices in the TDE hosts are driven by a central point source. Bright nuclei are also seen in post-starburst galaxies, and \citet{Yang2008} find that most post-starburst galaxies have lower \Sersic indices after masking the central 3 pixels.

\subsection{Galaxy Color Gradients}
\label{sec:colorgradients}

\begin{figure}
\includegraphics[width = 0.5\textwidth]{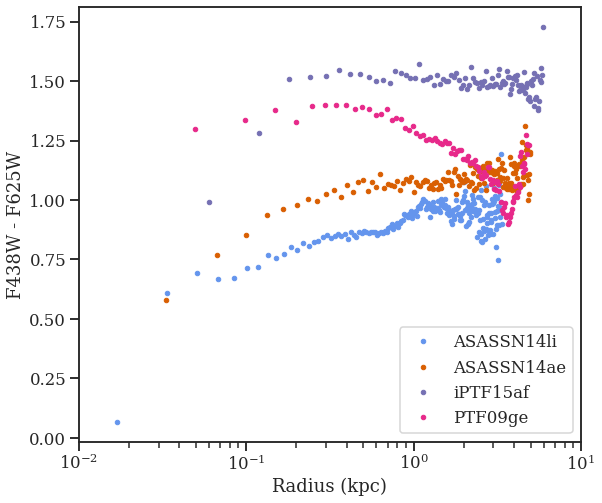}
\caption{Radial color profiles for the four TDE host galaxies. Three of the four (all except for the PTF09ge host) have blue color gradients, with bluer centers, consistent with other studies of post-starburst galaxies. The PTF09ge host has a flat color gradient in the center, with bluer outskirts due to a blue ring. The two galaxies best-fit with central point sources (ASASSN-14li and PTF09ge hosts) show different behaviors, with the ASASSN-14li host having a bluer center than the rest of the galaxy, and the PTF09ge host having a center with similar color to the rest of the galaxy. The blue dip in the color gradient of the PTF09ge host galaxy at $\sim3$ kpc is due to the star-forming ring that can be seen in Figure \ref{fig:color}.
}
\label{fig:colorgradients}
\end{figure}

We consider the TDE host galaxy color gradients in Figure \ref{fig:colorgradients}. We use the F438W and F625W band images, and the best-fit centers and sky levels, to determine the B-R color radial profiles. Three of the four TDE host galaxies (all except for PTF09ge) have blue color gradients, such that their centers are bluer than their outskirts. Blue color profiles are often seen in post-starburst galaxies \citep{Yang2008} because of the centrally-concentrated young/intermediate stellar populations, but are the reverse of the red central bulge - blue outer disk profiles often seen in star-forming galaxies and the flat or slightly red profiles seen in early type galaxies \citep{Wise1996}.

The blue color gradient (bluer at the center and redder towards higher radii) in the host galaxy of ASASSN-14li is consistent with those seen in most post-starburst galaxies. The quiescent Balmer-strong TDE host galaxies also have blue color gradients. Early type galaxies have a narrower range of color gradients than post-starburst galaxies, with most having flat or slightly red color gradients \citep{Yang2008}. Only the host galaxy of PTF09ge has a red color gradient, consistent with the early type sample. Even though not all of the TDE host galaxies are post-starburst, their color gradient distribution is consistent with the \citet{Yang2008} study of post-starburst galaxies, where 60\% had blue color gradients, 20\% had red color gradients, and 20\% had flat color gradients.

In order to quantitatively compare the color gradients to the post-starburst galaxies studied by \citet{Yang2008}, we fit both single and double power law slopes to the TDE host B-R color gradients. The double power law slopes allow us to separately parameterize the inner and outer slopes, using a break radius that is allowed to vary. Using either a Kolmogorov-Smirnov or Anderson-Darling test, we cannot rule out the possibility that the color gradient slopes, inner slopes, or outer slopes for the TDE host galaxies and post-starbursts are drawn from the same distribution.

The two galaxies best-fit with central point sources (ASASSN-14li and PTF09ge) show different behaviors, with ASASSN-14li having a blue central point source, and PTF09ge having a point source with similar color to the rest of the galaxy. These colors may point at different physical origins for the central point sources. While a blue central point source could be caused by a central star cluster, a concentrated stellar light profile, or an AGN, a red central point source could be a concentrated older stellar population or a dust-obscured blue source. In the case of the PTF09ge host galaxy, however, we observe no dust lanes near the core, and indeed a TDE was observed, limiting the possibilities for heavy nuclear dust extinction.

\subsection{Stellar Mass Surface Densities}
\label{sec:smsd}

Because bright central surface brightnesses could be due to either high stellar densities or low mass to light (M/L) ratios from younger stellar populations, we consider next the stellar mass surface densities of the TDE host galaxies. We use the F438W-F625W color profiles of the TDE host galaxies and the relations from \citet{Bell2001} to estimate the M/L profiles. We use a rolling median of the radial color profiles, smoothed over 4 pixels to reduce noise and account for the width of the PSF. We use the SDSS spectra and Keck spectra (PTF09ge host) used to initially classify the galaxies in \citet{French2016} to convert the observed \hst colors to rest-frame B-R colors. \citet{Bell2001} find that uncertainties in color-based M/L values are 0.1--0.2 dex, except in the case of a recent $\sim10$\% starburst, where uncertainties are $\sim0.25$ dex. We adopt a 0.25 dex uncertainty for the post-starburst host of ASASSN-14li, and 0.15 dex uncertainty for the other three host galaxies. The uncertainty in the M/L ratio dominates the error in the central stellar mass surface densities. We assume all the light is stellar in origin, with no contribution from a central persistent AGN.

For the comparison early type galaxy sample, we use the M/L values from \citet{Cappellari2013b}. Because the early type galaxies have less radial variation in color, we assume these M/L values for the entire radial extent of the early type galaxies. 

\begin{figure*}
\includegraphics[width = 0.5\textwidth]{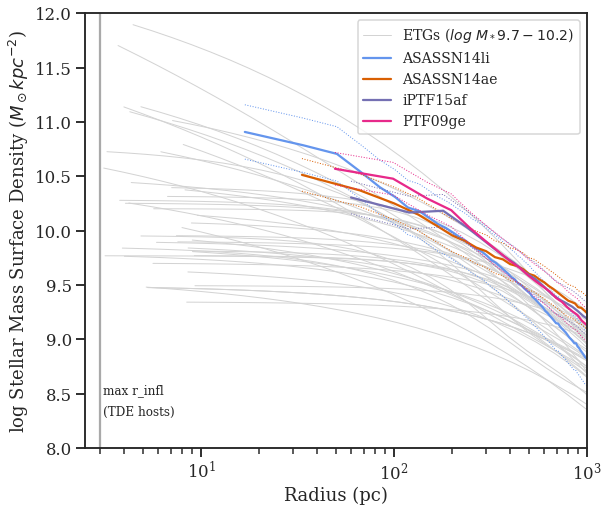}
\includegraphics[width = 0.5\textwidth]{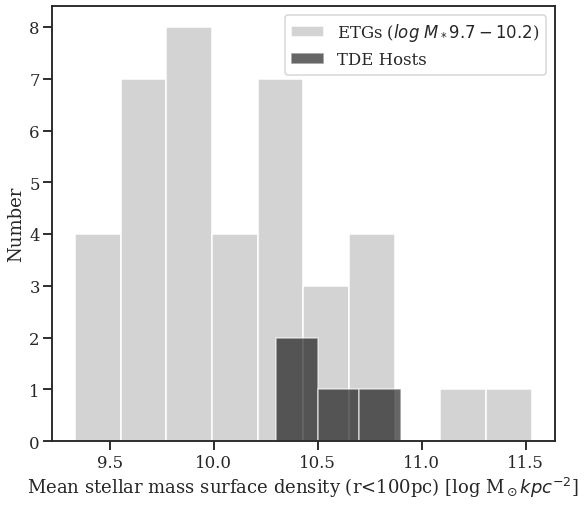}
\caption{{\it Left:} Stellar Mass Surface Density profiles for TDE host galaxies and comparison early type galaxies from Atlas-3D \citep{Krajnovic2013,Cappellari2013b}, derived from the surface brightness profiles (Figure \ref{fig:profiles}) and color profiles (Figure \ref{fig:colorgradients}). We consider the best-fit \Sersic+ disk models for the TDE hosts as in Figure \ref{fig:profiles}. Dotted lines indicate the range of uncertainties for the TDE host galaxies due to the uncertainty in the M/L ratios. If we assume that all of the central light from the TDE host galaxies is from stars, even after correcting for the lower M/L from the younger stars, the TDE host galaxies have higher stellar mass densities in their central 30--100 pc than most early type galaxies with similar total stellar mass. 
{\it Right:} Histograms of the mean stellar mass surface density within 100pc for the TDE host galaxies and the same comparison early type galaxies. The TDE host galaxies have stellar mass surface densities on these scales higher than 2/3 of the early type galaxies with similar total stellar mass. 
}
\label{fig:smsd}
\end{figure*}

Using the M/L profiles derived from the color gradients and the best-fit \Sersic+ disk models for the TDE hosts, we calculate the stellar mass surface density profiles. We plot the stellar mass surface density profiles in Figure \ref{fig:smsd}a. If we assume all of the central light from the TDE host galaxies is from stars, even after correcting for the lower M/L of the younger stars, the TDE host galaxies have higher stellar mass densities in their central 30--100 pc than most early type galaxies with similar total stellar mass. 

We note that these results do not depend on the PSF modelling method, and we find the stellar mass surface densities to not change more than the uncertainties if we vary our PSF modelling assumptions or do not PSF-correct the models. 

We compare the mean stellar mass surface densities within 100pc for the TDE host galaxies and the comparison early type galaxies (Figure \ref{fig:smsd}b). The TDE host galaxies have stellar mass surface densities on these scales higher than 2/3 of the early type galaxies with similar total stellar mass. We discuss the implications for how this increased central stellar density could raise the TDE rate in these galaxies in \S\ref{sec:discussion_density}.

\subsection{Quantitative Asymmetries}
\label{sec:asym}

Recent major or minor mergers will result in tidal features and disturbed morphologies in galaxies on large $\sim$kpc scales, which can be quantified through the asymmetric components of the stellar light \citep{Conselice2000}. Using a galaxy image rotated 180 degrees as a model for the symmetric stellar light, the residual light can reveal asymmetric features indicative of a recent merger.

We quantify the presence of asymmetric components using the asymmetry measure defined by \citet{Conselice2000}:
\begin{equation}
    A = \text{min} \Big (\frac{\Sigma |I_0 - I_{180}|}{\Sigma |I_0|} \Big ) - \frac{\Sigma |B_0 - B_{180}|}{\Sigma |I_0|}, 
\end{equation}
where $I$ is the flux of the image $(I_0)$ and the image rotated 180 degrees $(I_{180})$. Each sum is over a circular aperture of radius $2\times$ the Petrosian radius. $B_0$ represents background pixels and their rotated image ($B_{180}$) summed over the same size aperture. Because the galaxy centers may not be at the exact center of our pixels, we iterate the rotation center in a small 7$\times$7 pixel grid, and choose the center in which the sum of the squared asymmetric residuals is the least.

We compare this value to the concentration
\begin{equation}
    C = 5 \text{log} \Big (\frac{r_{80}}{r_{20}} \Big )
\end{equation}
where $r_{80}$ and $r_{20}$ and the radii containing 80\% and 20\% of the flux, respectively, as in \citet{Conselice2003a}. The C--A plane separates out early type, late type, and starburst galaxies due to the presence of central bulges, spiral arms, and merger companions. \citet{Yang2008} found post-starburst galaxies to lie at slightly higher concentrations and much higher asymmetries than early type galaxies.

\begin{figure*}
\includegraphics[width=\textwidth]{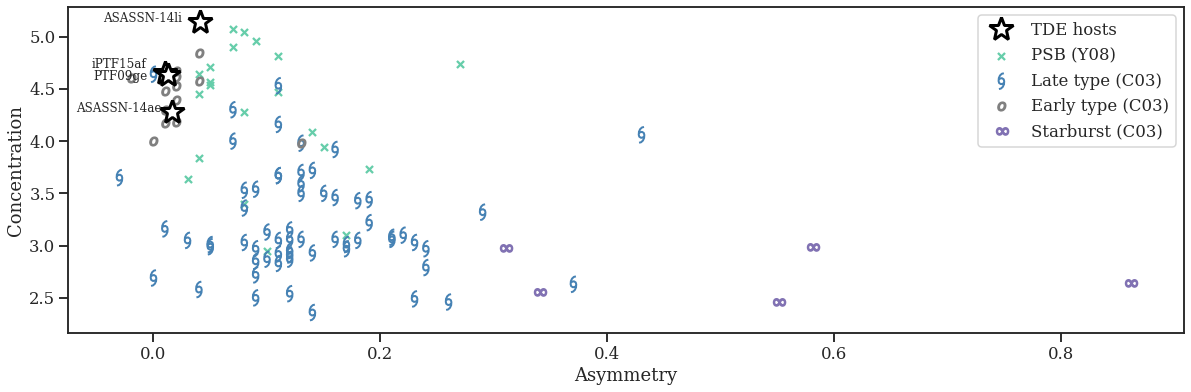}
\caption{Concentration vs. Asymmetry for TDE host galaxies and comparison late type, early type, and starburst galaxies from \citet{Conselice2003a} and comparison post-starburst galaxies from \citet{Yang2008}. \citet{Conselice2003a} found this plane to separate late type galaxies from early type galaxies, and to further separate out starburst galaxies from late/early types. \citet{Yang2008} found post-starburst galaxies to lie in a unique space with slightly higher concentrations than early types and with higher asymmetries. The four TDE host galaxies considered here have high concentrations like the post-starburst and early type galaxies, but low asymmetries like the early types. The host galaxy of ASASSN-14li has the highest asymmetry, yet is similar to the lowest asymmetry post-starburst galaxies. This is consistent with the qualitative impression from the previous figures showing a lack of obvious strong tidal features in the TDE hosts like what has been seen for $\sim$half of post-starburst galaxies \citep{Yang2008}. The concentration parameter here is a global (kpc scale) measure, and not sensitive to the innermost slope of the host galaxies as studied elsewhere in this paper.}
\label{fig:conc_asym}
\end{figure*}

We compute $C$ and $A$ for the TDE host galaxies and compare it to early type, late type, and starburst galaxies from \citet{Conselice2003a} and post-starburst galaxies from \citet{Yang2008} in Figure \ref{fig:conc_asym}. 

We use the F625W band to be most consistent with the post-starburst analysis. The \citet{Yang2008} sample of post-starburst galaxies is at slightly higher redshift ($\langle z \rangle = 0.09$ vs. $\langle z \rangle = 0.05$) than the TDE host galaxies considered here, and the observations of the TDE host galaxies are deeper, at $26.3-27$ mag$_{\text{AB}}$ arcsec$^{-2}$ vs. $25.3$ mag$_{\text{AB}}$ arcsec$^{-2}$ for the \citet{Yang2008}. Thus, the TDE host galaxy observations presented here should be at least as sensitive to faint tidal features as \citet{Yang2008}. 

Tidal features will fade with time after the merger, but the \citet{Yang2008} sample has a similar post-starburst age distribution as the TDE host galaxies. The \citet{Zabludoff1996} method used to select the sample used by \citet{Yang2008} selects similar galaxies as the age-dated sample considered by \citet{French2018b}. \citet{French2017} test whether the TDE host galaxies have a distinguishable distribution of post-burst ages compared to the post-starburst galaxies selected from the SDSS, and find no statistically significant difference. Thus, it is unlikely that the lower asymmetries seen in the TDE hosts compared to the \citet{Yang2008} sample of post-starburst galaxies is due to older post-merger ages in the TDE host galaxies.

The imaging data used by \citet{Conselice2003a} is from the \citet{Frei1996} sample of nearby galaxies. Because the physical spatial resolution and signal to noise ratio in each of the datasets considered here are high, the uncertainty in concentration and asymmetry due to variations in data quality will be low. Using the tests done by \citet{Lotz2004}, \citet{Yang2008} expect $\Delta A \lesssim 0.05$ and $\Delta C \lesssim 0.1$ if the SNR $\gtrsim 7$ and resolution $\lesssim 100$ pc/pixel. These criteria are also satisfied by the TDE host galaxy imaging, so the values of A and C can be directly compared between each of the samples considered here.

We calculate the asymmetry within $2\times$ the Petrosian radius to be consistent with \citet{Yang2008}, but we note that if we used 1.5$\times$ as done by \citet{Conselice2003a}, the asymmetries would be higher by only 0.002--0.004.  

The four TDE host galaxies considered here have high concentrations like the post-starburst and early type galaxies, but low asymmetries like the early types. Low asymmetries in TDE host galaxies have also been noted by \citet{Law-Smith2017} using SDSS imaging, though we note that that analysis uses residual asymmetries, where smooth models are first subtracted from the data before rotating.

\begin{figure*}
\includegraphics[width = 1\textwidth]{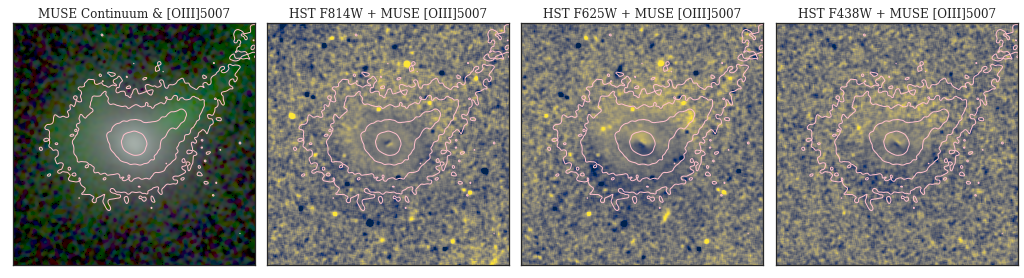}
\caption{
The host galaxy of ASASSN-14li has large asymmetric features visible in ionized emission from extended gas around the galaxy, revealed by MUSE observations \citep{Prieto2016}. In the left panel, we show a three-color image with red and blue continuum regions from the MUSE datacube and green from the [OIII]5007 line. Pink contours for the flux at rest-frame 5007\AA\ are shown in each panel. In the right three panels, we compare the extended ionized feature to the asymmetric residuals from each band of \hst imaging (see \S\ref{sec:asym}). Each panel is 11.2\arcsec$\times$11.2\arcsec. A faint residual feature seen to the upper right of the galaxy in the F625W image is aligned with the extended [OIII]5007 arm. This feature is not seen in the F814W or F438 images. Thus, the feature in the F625W image may not be due to stellar light, but instead H$\alpha$ and/or [NII]6584 emission from extended ionized gas, as these lines are in the F625W band at the redshift of the ASASSN-14li host galaxy. However, we note that the F625W image is deeper than the other two images by $\sim0.3$ mag, so the observed feature could be faint stellar light. The sum of the F625W residuals nonetheless results in less asymmetry than most other post-starburst galaxies. Thus, the recent starburst in this galaxy may have been triggered by a more minor $>3:1$ merger than typical for other post-starburst galaxies. The two SMBHs in each merger pair may not have had time to coalesce via dynamical friction to a bound SMBHB. 
}
\label{fig:musecomp}
\end{figure*}

The host galaxy of ASASSN-14li has the highest asymmetry of the TDE hosts, yet is similar to the lowest asymmetry post-starburst galaxies, with 80\% of the \citet{Yang2008} post-starburst galaxies having higher asymmetries. In contrast to the relatively undisturbed morphology seen in the stellar light, this galaxy has extended regions of ionized gas observed using MUSE \citep{Prieto2016}. Extended ionized regions may be caused by past AGN activity and have been seen around other galaxies in large imaging surveys  \citep[``voorwerps"][]{Lintott2009, Keel2017}. We compare the morphology of the extended [OIII]5007 bright filaments with the asymmetric residuals in Figure \ref{fig:musecomp} to search for corresponding features in the stellar light. A faint residual feature seen to the upper right of the galaxy in the F625W image is aligned with the extended [OIII]5007 arm. This feature is not seen in the F814W or F438 images. Thus, the feature in the F625W image may not be due to stellar light, but instead H$\alpha$ and/or [NII]6584 emission from extended ionized gas, as these lines are in the F625W band at the redshift of the ASASSN-14li host galaxy. However, the F625W image is deeper than the other two images by $\sim0.3$ mag, so the observed feature could be faint stellar light. The sum of the F625W residuals nonetheless results in an asymmetry below most other post-starburst galaxies. Thus, the recent starburst in this galaxy may have been triggered by a more minor merger than typical for other post-starburst galaxies. We explore this possibility in more detail in \S\ref{sec:smbhb}.

\begin{figure}
\includegraphics[width = 0.5\textwidth]{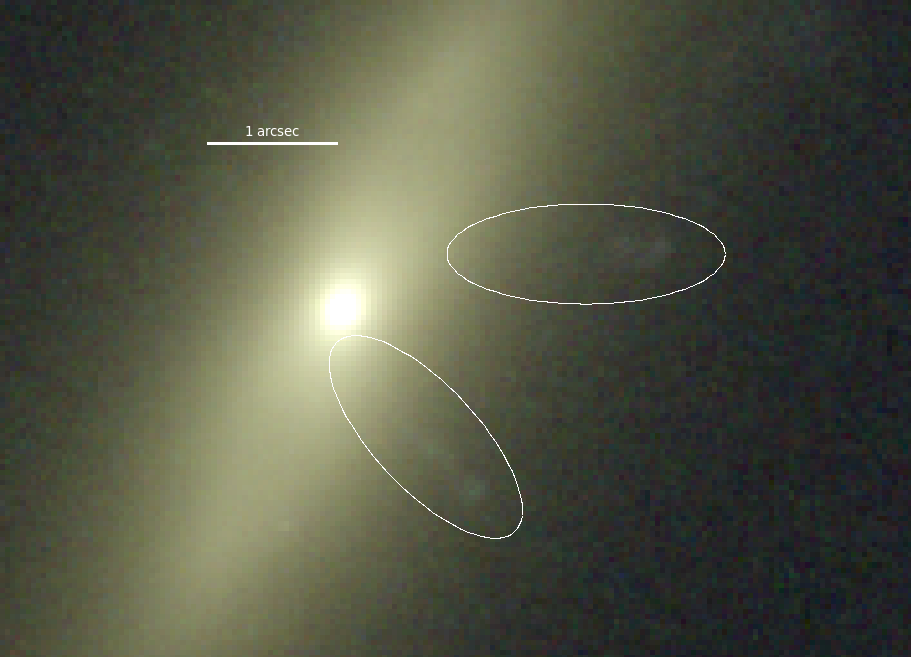}
\caption{Three color image of the ASASSN-14ae host galaxy, with asymmetric features circled. This light may be stellar light from a recent merger, or ionized gas from AGN activity or star formation, as it is similar to the outlines of an outflow cone. Integral field spectroscopy will be required to test the nature of these asymmetric features. The overall asymmetry is not larger than typical of early type galaxies (Figure \ref{fig:conc_asym}).
}
\label{fig:14ae}
\end{figure}

The host galaxies of ASASSN-14ae and iPTF15af are quiescent Balmer-strong, and have asymmetries consistent with early type (quiescent) galaxies. The only features that contribute to their low asymmetries are two faint features extending from the nucleus of the host galaxy of ASASSN-14ae (Figure \ref{fig:14ae}) and the blue clumps surrounding the host galaxy of iPTF15af (which can be seen in Figure \ref{fig:color}). Despite these features, the asymmetries we calculate are consistent with early type galaxies. In the absence of a {\it HST}--observed sample of quiescent Balmer-strong galaxies, we expect them to be similar to both early type and post-starburst galaxies in C--A space, as they have similar post-burst ages as the post-starburst sample, with generally lower fractions of stellar mass formed in the recent starbursts \citep{French2017}. The lack of strong asymmetries in the two quiescent Balmer-strong TDE hosts may indicate their recent bursts of star formation were triggered by minor mergers, other interactions, or internal instabilities that were sufficient to drive gas to their centers and form stars, but not enough to create extended tidal features typical after major mergers. 

The host galaxy of PTF09ge was classified as quiescent, although the blue outer ring would have been partially missed by the SDSS fiber used in classification, and likely has more star formation. The extended arms seen in Figures \ref{fig:color} and \ref{fig:stretch} are nearly symmetric with one another. Despite the blue star-forming ring, the host galaxy of PTF09ge has an asymmetry and concentration more similar to early type galaxies than star-forming galaxies. 

We consider the implications of the observed morphologies on the recent merger histories and possibilities for rapidly producing a bound SMBHB in \S\ref{sec:smbhb}.

\subsection{Elliptical Isophote Fitting}
\label{sec:ellipse}

\begin{figure*}
\centering
\includegraphics[width = 0.8\textwidth]{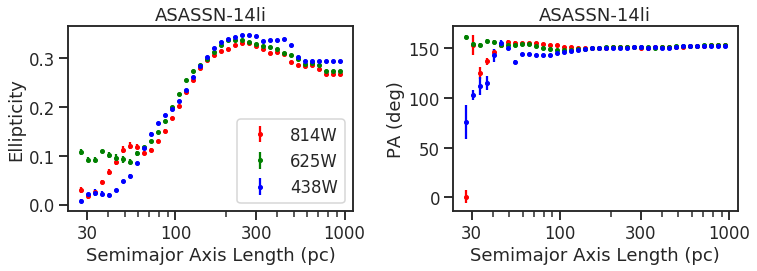}
\includegraphics[width = 0.8\textwidth]{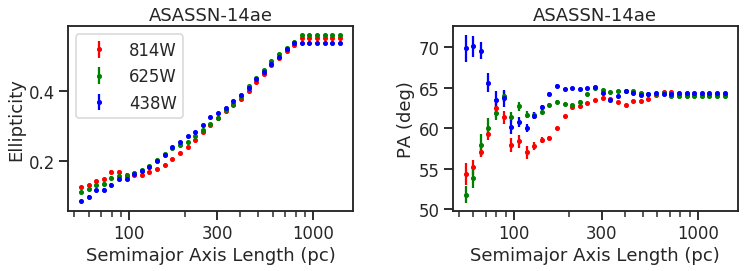}
\includegraphics[width = 0.8\textwidth]{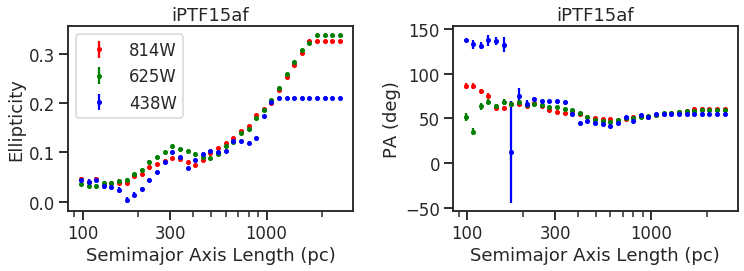}
\includegraphics[width = 0.8\textwidth]{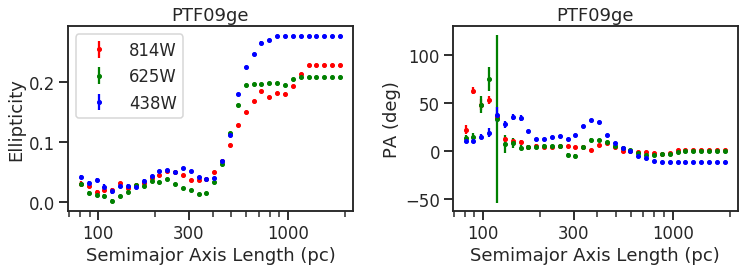}
\caption{Results of elliptical isophotal fitting for the four TDE host galaxies (for all three \hst photometric bands). The semi-major axis starts at 1.5 pixels and increments in steps of $1.1\times$. All four TDE host galaxies have ellipticities that decrease towards the galaxy centers. We compare the central ($<140$ pc) ellipticities to early type galaxies in Figure \ref{fig:ellipse_comp}. The radial trends in position angle vary, with shifts of 10--100 degrees seen over $\sim100$ pc changes in radius. The bluest band (F438W) is significantly affected by the star-forming components of the galaxy, and the green band (F625W) may be altered by the presence of H$\alpha$ and NII line emission.
}
\label{fig:ellipse}
\end{figure*}

The shapes of the central isophotes of the TDE host galaxies reveal information about the distribution of nuclear stellar orbits. In order to further quantitatively explore the host galaxy morphologies, we use the {\tt astropy} package {\tt photutils} \citep{larry_bradley_2019_2533376} to perform elliptical isophotal fitting on the TDE host galaxies. The resulting best-fit ellipticity and position angles as a function of semi-major axis are shown in Figure \ref{fig:ellipse}. The radial trends in position angle vary, with shifts of 10--100 degrees seen over $\sim100$ pc changes in radius. The ellipticities do not appear to vary with wavelength, with the exception of the F625W band at the center of the ASASSN-14li host galaxy. Due to the presence of H$\alpha$ and NII lines in this band, this measurement may also include non-stellar light. There are several discrepancies by band in the position angle measurements, as the F438W band traces the different features from on-going star formation and its associated bluer light, especially for the host galaxy of PTF09ge.

\begin{figure}
\includegraphics[width = 0.5\textwidth]{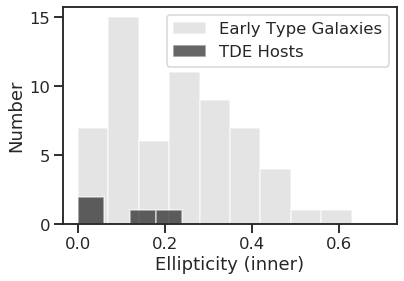}
\caption{Comparison of inner ellipticities from elliptical isophote fitting for early type galaxies \citep[F555W filter]{Lauer2005} and the TDE host galaxies (F625W filter). \citet{Lauer2005} calculate the inner ellipticities for the early type galaxies using the luminosity-weighted mean over the central region inside the break radius; for the TDE host galaxies, we calculate the inner ellipticities inside the region with radius $<140$ pc (the mean break radius), due to the lack of good Nuker profile fits to the TDE host galaxies.  The TDE host galaxies have relatively small inner ellipticities compared to the early type sample. These ellipticities are smaller than the $e\sim0.5$ where stars at the inner disk will start to be lost to the SMBH at an increased rate \citep{Madigan2018}. The low central ellipticities suggest an absence of strong radial anisotropy in the nuclear stellar orbits, though we cannot rule out eccentric disks at the scale of the black hole gravitational radius of influence ($r\sim0.3-3$ pc). 
}
\label{fig:ellipse_comp}
\end{figure}

We compare the central ellipticities of the TDE host galaxies to those obtained by \citet{Lauer2005} for the Nukers sample of early type galaxies using similar methods. We use the F625W filter images for the TDE hosts to be most consistent with the F555W filter images used by \citet{Lauer2005}. While \citet{Lauer2005} calculate the inner ellipticities for the early type galaxies using the luminosity-weighted mean over the central region inside the break radius, we calculate the inner ellipticities for the TDE host galaxies inside the region with radius $<140$ pc (the mean break radius), due to the lack of good Nuker profile fits to the TDE host galaxies. The inner ellipticities for the TDE host galaxies and early type galaxies are shown in Figure \ref{fig:ellipse_comp}. 

The TDE host galaxies have relatively small inner ellipticities compared to the early type sample. These ellipticities are smaller than $e\sim0.5$ where stars at the inner disk will start to be lost to the SMBH at an increased rate \citep{Madigan2018}, though we cannot rule out an eccentric disk at smaller scales closer to the radius of influence $r\sim0.3-3$ pc.

While low inner ellipticities can be related to high central concentrations in galaxies (due to the formation of stars after gaseous dissipation in a gas-rich merger \citealt{Cox2006}), the low inner ellipticities represent additional information about the galaxy centers. Post-merger early type galaxies with extra central light in their surface brightness profiles are more likely to have lower ellipticities, but with significant scatter \citep{Hopkins2009}. A central unresolved point source component could also contribute to a low central ellipticity, but the PSF scale is only FWHM$\sim2$ pixels, and the low central ellipticities in the two TDE host galaxies best-fit with central point source components in \S\ref{sec:sb_profiles} (ASASSN-14li and PTF09ge) extend to major axis lengths of $\sim8$ pixels.

We consider the impact of these observations on the various dynamical effects that could increase the TDE rate in \S\ref{sec:dyn}.

\section{Implications for the TDE rate per galaxy}

Several different hypotheses have been proposed to explain the enhanced rate of TDE production in post-starburst and quiescent Balmer-strong galaxies.  These include very dense central star clusters \citep{Stone2015}, SMBHBs \citep{Arcavi2014}, strongly asymmetric geometries in the central potential \citep{French2016}, anisotropic nuclear stellar populations \citep{Stone2018}, and secular instabilities in nuclear stellar disks \citep{Madigan2018}.  With so many plausible explanations, it is important to find observable tests to discriminate among different dynamical scenarios.

Past studies of TDE host galaxies have examined (i) the distribution of post-starburst ages and star formation histories \citep{French2016, French2017}; (ii) broader structural properties of TDE host galaxies visible with low-resolution photometry \citep{Law-Smith2017, Graur2018}; and (iii) inferred SMBH masses in TDE populations \citep{Wevers2017, Mockler2018, Wevers2019}.  Of these observations, post-starburst ages can be compared to theoretical delay time distributions, and SMBH mass inferences can be compared to theoretical rate predictions as a function of host mass \citep{Stone2018}.  Here we discuss the ways in which our {\it HST}-resolution host galaxy imaging can constrain different theoretical scenarios.

\subsection{High Central Stellar Densities}
\label{sec:discussion_density}
In a simple galactic nucleus (i.e., a quasi-spherical star cluster surrounding a single SMBH), TDE rates are set by diffusion of stars into a phase space loss cone.  Diffusion of stellar orbital parameters is driven by two-body scatterings, the rate of which scales with stellar density.  Holding all else equal, the TDE rate will therefore increase if the stellar density increases. Because the bulk of the loss cone flux comes from near the black hole gravitational influence radius ($r_{\rm infl}\sim0.3-3$ pc for the black hole masses of observed TDEs), galaxies with a density enhancement on this scale are expected to have a high TDE rate. 

Using SDSS imaging and multiple parameterizations of stellar concentration, TDE host galaxies have been observed to have high central surface brightnesses on kpc scales \citep{Law-Smith2017, Graur2018}. While SDSS imaging alone cannot resolve scales near $r_{\rm infl}$ in distant TDE hosts, if one extrapolates kpc-scale properties inwards, the dependence of TDE rates on concentration found by \citet{Graur2018} is consistent with theoretical predictions by \citet{WangMerritt2004}. {\it HST} observations of the nearby post-starburst galaxy NGC 3156 find an unusually steep central surface brightness profile on scales close to $r_{\rm infl}$\footnote{The pixel scale of the NGC3156 observations studied by \citet{Stone2016} is 4.3 pc, and the expected influence radius is $r_{\rm infl} \approx 3$ pc.}; loss cone modeling implies an elevated TDE rate matching that observed in post-starburst galaxies \citep{Stone2016}. Simulations also show an increased TDE rate after a galaxy-galaxy merger due to stellar overdensities caused by the merger \citep{Pfister2019}. 

In this work, we have shrunk to $30-100$ pc the spatial scales at which we observe TDE host galaxies to have high surface brightnesses. These scales are still above $r_{\rm infl}$, but nevertheless demonstrate that the unusually high surface brightness seen on $\sim$kpc scales (in ground-based imaging) are correlated with smaller-scale high concentration effects. In our analysis of the central stellar surface densities (\S\ref{sec:smsd}), we also find that---in comparison to early type galaxies of similar mass---the TDE host galaxies have high stellar surface densities on $30-100$ pc scales, using the TDE host color gradients to account for variations in M/L with radius. The radial color gradients identify a bluer, younger central population in three out of four of the TDE host galaxies we have studied (Fig. \ref{fig:colorgradients}), providing further evidence that it was the past star formation episode that created an unusually dense central stellar population.  While there are not yet post-starburst TDE hosts close enough for {\it HST} to resolve down to $r_{\rm infl}$, this result is consistent with {\it HST} observations of the nearby post-starburst NGC 3156 \citep{Stone2016}. Thus, it is likely that the TDE host galaxies have a greater number of stars available in orbits where a star could be tidally disrupted. 

\subsection{Black Hole Binaries}
\label{sec:smbhb}
The TDE rate can be greatly enhanced during the late inspiral phase of a supermassive black hole binary (SMBHB) \citep[e.g.,][]{Ivanov2005, Chen2011}, as the two black holes reach the nucleus and form a bound pair. Because mergers can both produce a SMBHB and trigger a starburst, establishing whether the TDE host galaxies have experienced a recent merger (major or minor) is important for assessing whether the TDE rate is impacted by SMBHB effects. 

The recent merger history of galaxies can be probed by the presence of tidal features and other asymmetries. We explore the asymmetric features of the TDE host galaxies in \S\ref{sec:asym} and compare them to normal early type and star forming galaxies, starbursts, and post-starbursts. While high asymmetries caused by merger features are common in post-starburst galaxies \citep{Yang2008}, the one post-starburst TDE host considered here (ASASSN-14li) is consistent with only the low asymmetry tail of the post-starburst distribution. The other three TDE host galaxies show even less asymmetry, consistent with early type galaxies. 

A recent gas-rich major merger (with mass ratio $\sim$ 1:1--1:3) consistent with producing a post-starburst signature \citep{Snyder2011} will produce tidal features with $A>0.35$ lasting for hundreds of Myr \citep{Lotz2010} before and during the starburst phase, and elevated above the level of early type galaxies throughout the post-starburst phase \citep{Yang2008}. The lack of high asymmetries seen in the TDE hosts considered here implies they have not experienced recent gas-rich major ($\sim$ 1:1--1:3) mergers. Indeed, the low mass fractions produced in the recent starburst for the quiescent Balmer-strong hosts are more consistent with being triggered by a more minor merger, and even the extended ionized features seen in the host galaxy of ASASSN-14li could be caused by a $\sim$ 1:3--1:5 mass ratio merger since even minor mergers can create tidal debris \citep{Tal2009}. 

The mass ratio of the recent merger is important in the context of the TDE rate, because it sets the dynamical friction timescale from the start of the merger to the formation of a bound SMBHB. A starburst episode is likely to be triggered upon the first passage of the two galaxies and at their final coalescence \citep{Mihos1994,Cox2008,Renaud2015}. The dynamical friction timescale from N-body simulations is $\sim1.4$ Gyr for a 1:1 mass ratio major merger\footnote{The dynamical friction timescale for an equal mass merger trends towards the dynamical timescale of $\sim10$\% of the Hubble time. We consider here only $z\sim0$ TDEs, but note that these timescales are faster at higher redshift, and indeed at $z\sim3$ more minor mergers can form bound SMBHBs more quickly \citep[e.g.,][]{VanWassenhove2014,Khan2018}.}, and scales with the mass ratio such that the dynamical friction timescale will be $\sim2\times$ longer for a 1:3 merger and $\sim3\times$ longer for a 1:5 merger \citep{Boylan-Kolchin2008}. Thus, even allowing for a Gyr separation between the start of the merger and the triggering of the starburst, unequal mass mergers are unlikely to have formed a SMBHB in the 500 Myr required by the post-burst ages \citep{French2017} of the three post-starburst/quiescent Balmer-strong host galaxies.

The lack of strong merger features can, in principle, allow us to rule out the possibility of a major merger recent enough to have formed a SMBHB capable of a high TDE rate.  A detailed morphological comparison of our four galaxies with predictions (from cosmological simulations) for the duration of post-merger features, the inspiral time to form a SMBHB, and the evolution of the stellar populations is beyond the scope of this paper. 

To summarize, we have found that the post-starburst and quiescent Balmer-strong nature of these TDE host galaxies does not generally arise from recent major ($\sim$1:1 mass ratio) mergers. This result  qualitatively disfavors the hypothesis that elevated TDE rates are due to SMBHBs and may be quantifiable in the future through morphological comparisons to galaxy simulations.

\subsection{Dynamical Effects}
\label{sec:dyn}
A variety of other effects relating to the dynamics of the central stars have been proposed to increase the TDE rate, such as a radial anisotropy \citep{Stone2018} or an eccentric disk \citep{Madigan2018}. As discussed in \S\ref{sec:ellipse}, we observe low central ellipticities in the TDE host galaxies, but the resolution of these observations is not sufficient to observe the scale of the eccentric disk ($\sim r_{\rm infl}$) proposed by \citet{Madigan2018}. 

On larger scales, an excess of stars with eccentric orbits can still contribute to a high TDE rate through the radial anisotropy scenario considered by \citet{Stone2018}.  Velocity anisotropy is often quantified with the $\beta$ parameter, which ranges from purely tangential orbits ($\beta=-\infty$) to purely radial orbits ($\beta=1$); an isotropic galaxy model has $\beta=0$.  Because mild radial velocity biases wash out (due to relaxation) fairly quickly, nuclear starbursts must create a large initial bias ($\beta \gtrsim 0.5$) in order to explain the observed rate enhancement in post-starburst galaxies \citep{Stone2018}. While this mechanism can operate even in a quasi-spherical geometry, the requisite $\beta$ values are so large as to be near the onset for the radial orbit instability \citep{Henon1973}.  The nonlinear outcome of this instability is to convert initially spherical geometries into highly flattened ones \citep{PolyachenkoShukman1981} that may be nearly prolate or (for the largest $\beta$ parameters) significantly triaxial \citep{MerrittAguilar1985}.  

Although most TDEs are sourced from scales of $r_{\rm infl}$ in a quasi-isotropic galactic nucleus, this is not the case in a radially biased--galaxy; for large $\beta$ values, the peak of the loss cone flux can be up to an order of magnitude beyond $r_{\rm infl}$, a distance that may be resolved for the host of ASASSN-14li (and is only unresolved by a factor of a few for our other galaxies). It is therefore significant that we observe unusually low values of nuclear ellipticity in the centers of all TDE host galaxies. If elevated TDE rates in post-starburst galaxies were due to the production of $\beta\gtrsim 0.5$ anisotropies in nuclear starbursts, many TDE host galaxies should undergo the radial orbit instability and acquire prolate-to-triaxial nuclear geometries.  The absence of evidence for this in four TDE host galaxies is a significant indication that radial orbit biases do not explain the TDE host galaxy preference.

The low central ellipticities we observe are not inconsistent with triaxial geometries, however. A central bar (on $<100$ pc scales) or many other non-axisymmetric central geometries\footnote{We note, however, that self-similar triaxial ellipsoids are one type of non-axisymmetric geometry that is consistent with our observations.} can elevate TDE rates by an order of magnitude \citep{Merritt2004}. Triaxialities can be observed through changes in the position angle with radius. We examine the position angles in \S\ref{sec:ellipse}, finding a range of variations in the PA with radius at higher radii ($>$ 100 pc). However, we do not have the resolution to be sensitive to position angle twists on scales $<<$ 100 pc, as studied by \citep{Merritt2004}. Triaxiality will increase the TDE rate on scales smaller than the radius of gravitational influence \citep{Vasiliev2014}, to which our observations are not sensitive.

\setcounter{footnote}{0}

The observations of the surface brightness profiles and ellipticity profiles also allow us to test the possibility proposed by \citet{Cen2020} in which a massive $\sim$100 pc disk with $>$50\% of the stellar mass produces a high rate of TDEs around the secondary SMBH as it inspirals through the disk. There are two separate observational consequences, depending on whether the massive disk is closer to being edge-on or face-on. If the disk is closer to face-on, we would observe a dominant central disk component in our \galfit modelling (\S\ref{sec:sb_profiles}). However, the four TDE host galaxies considered here are better fit by a high \Sersic index model than a low \Sersic index or exponential disk\footnote{We note that the models we assume for the surface brightness profiles are observationally motivated, whereas the Mestel disk assumed by \citet{Cen2020} has a different form to facilitate the analytical calculations.}. If the massive disk were closer to edge on, we would observe elliptical isophotes in the central $\sim$100 pc, inconsistent with the low central ellipticities observed by the four TDE host galaxies considered here. However, the discovery and identification of TDEs around the secondary SMBHs considered by \citet{Cen2020} may be selected against as possible TDEs as they will be offset from the primary nucleus, and the primary SMBH could be an AGN. \citet{Cen2020} predict a high instance of repeating TDEs in such systems; studies of the host galaxies of any repeating TDEs would be greatly informative in constraining this possibility.

The presence of bars can also have an impact on the funnelling of gas toward the nuclei of galaxies and the formation of central eccentric disks \citep{Hopkins2010a,Hopkins2010b}, which may be related to the TDE rate \citep{Madigan2018, Wernke2019}. However, larger-scale bars such as those seen in the host galaxies of iPTF15af and potentially PTF09ge are common among low redshift galaxies \citep{Elmegreen2004}, and there is no evidence that barred galaxies have a higher TDE rate. Further study of the kinematics of TDE host galaxies on scales close to $r_{\rm infl}\sim0.3-3$ pc will be needed to determine the possible impact of these mechanisms. Depending on the SMBH mass, this will require a TDE within $\sim50$ Mpc (to resolve 10 pc scales) or within $\sim5$ Mpc (to resolve 1 pc scales).

\section{Conclusions}

The spectral properties of the host galaxies of many Tidal Disruption Events are unusual, showing evidence of a recent, but not on-going, burst of star formation in many cases. Here we explore the morphologies of four TDE hosts on 30 pc to $>$kpc scales with \hst WFC3 multi-band imaging. This sample includes one host that experienced a recent strong starburst (ASASSN-14li), two hosts with recent weaker starbursts (ASASSN-14ae and iPTF15af), and one early type galaxy (PTF09ge). We model their two-dimensional surface brightness distributions and compare them to those of post-starburst and early type galaxies in the literature. We summarize our conclusions here.

These TDE host galaxies not only have a range of star formation histories, but also are diverse morphologically, with some showing disk, bar, and ring components in addition to centrally concentrated stellar light profiles. Two of the four TDE host galaxies---ASASSN-14li and PTF09ge---have an unresolved central point source.

The TDE host galaxies have high central surface brightnesses on 30--100 pc scales, similar to post-starburst galaxies and with higher central surface brightnesses than early type galaxies of similar stellar mass.

Three of the four TDE host galaxies (except for PTF09ge) have blue color gradients, with bluer centers, similar to those of post-starburst galaxies. The two central point sources have different properties: ASASSN-14li's is blue, whereas PTF09ge's has a similar color to the rest of the galaxy. High resolution spectroscopy is required to determine the nature of these sources.

We use the color gradients to estimate the radial dependence of the mass-to-light ratio and thus the stellar mass surface density profiles. Compared to early type galaxies of similar stellar masses, the TDE hosts have high central (30--100 pc scales) stellar mass surface densities.

The disturbed kpc-scale morphology seen in roughly half of post-starburst galaxies \citep{Yang2008} is absent here. The morphological asymmetries for the TDE hosts are similar to those of early type galaxies, except for the post-starburst host galaxy of ASASSN-14li, which has an asymmetry characteristic of more relaxed post-starburst galaxies. The lack of strongly disrupted morphologies is inconsistent with recent major ($\sim$1:1 mass) mergers, although minor ($\lesssim$1:3) mergers are possible and could have triggered the starbursts. 

The three TDE hosts with recent starbursts span a range in the time since their bursts ended (from 150--600 Myr; \cite{French2017}), similar to the \citet{Yang2008} sample of post-starburst galaxies. Given these post-burst ages and merger mass ratio constraints, the dynamical friction timescales for the galaxies to coalesce after a minor merger \citep{Boylan-Kolchin2008} is longer than the post-burst ages, so the two central SMBHs may not yet be close enough to form a binary capable of boosting the TDE rate \citep[e.g.,][]{Ivanov2005, Chen2011}.

We perform elliptical isophote fitting to model the radial dependence of the ellipticity and isophote position angle. The TDE hosts have low central ($<140$ pc) ellipticities compared to early type galaxies. The low central ellipticities disfavor a strong radial anisotropy as the cause of the enhanced TDE rates, although we cannot rule out eccentric disks or nuclear triaxiality at the scale of the black hole gravitational radius of influence ($\sim1$ pc).

The high central surface brightnesses of these TDE host galaxies on scales of 30--100 pc and their lack of dramatic merger features on kpc scales suggest that the high stellar densities in the nucleus is more responsible in increasing TDE rates than the presence of SMBH binaries or radial anisotropies.

\acknowledgements
This work is based on observations made with the NASA/ESA Hubble Space Telescope, obtained at the Space Telescope Science Institute, which is operated by the Association of Universities for Research in Astronomy, Inc., under NASA contract NAS 5-26555. These observations are associated with program HST-GO-14717.001-A.  We thank Tod Lauer for useful discussions and insights into the data. 
K.D.F. is supported by Hubble Fellowship grant HST-HF2-51391.001-A, provided by NASA through a grant from the Space Telescope Science Institute (STScI), which is operated by the Association of Universities for Research in Astronomy, Inc., under NASA contract NAS5-26555. I.A. is a CIFAR Azrieli Global Scholar in the Gravity and the Extreme Universe Program and acknowledges support from that program, from the Israel Science Foundation (grant numbers 2108/18 and 2752/19), from the United States - Israel Binational Science Foundation (BSF), and from the Israeli Council for Higher Education Alon Fellowship.

This work made use of the SciPy \citep{jones_scipy_2001},  IPython \citep{PER-GRA:2007}, NumPy \citep{van2011numpy}, and Astropy \citep{2013A&A...558A..33A} packages.

\bibliographystyle{aasjournal}
\bibliography{extra_hostgal.bib}

\clearpage

\begin{appendix}
\section{High stretch Images}
\label{sec:resids}

\begin{figure*}
\centering
\includegraphics[width = 0.7\textwidth]{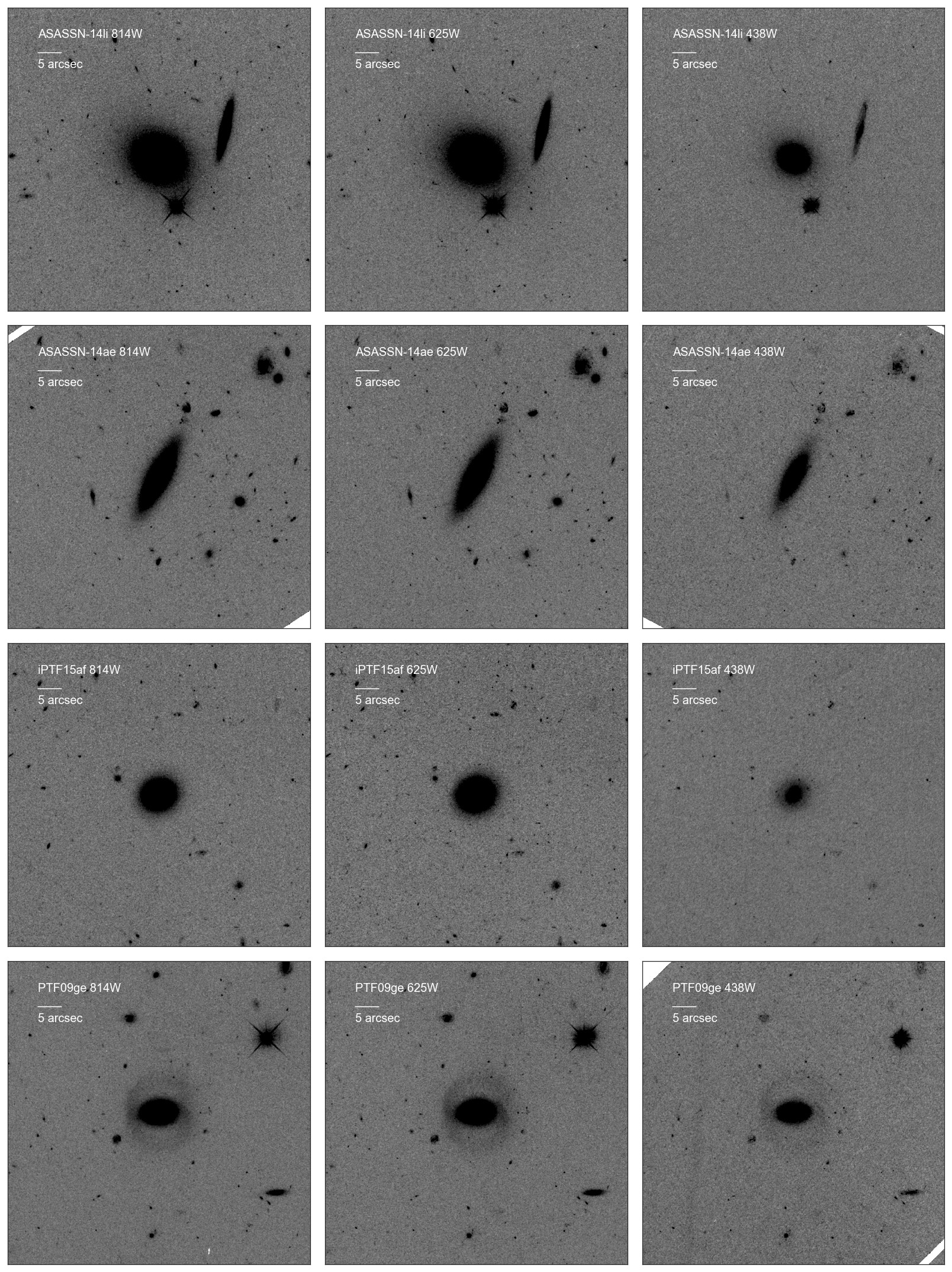}
\caption{High stretch images of the four TDE host galaxies in each of the three bands. This stretch is optimized to search for low surface brightness features. Two extended arms (possibly spiral arms or a ring-like ``$R_1^\prime$" feature \citep{Buta2017}) can be seen in the image of the PTF09ge host galaxy (bottom row).
}
\label{fig:stretch}
\end{figure*}

In Figure \ref{fig:stretch} we present high stretch images of the four TDE host galaxies in each of the three bands, optimized to search for low surface brightness tidal features.

\end{appendix}

\end{document}